\documentstyle[12pt,aaspp,psfig]{article}

\received{26 March 1998}
\accepted{19 May 1998}
\cpright{AAS}{1998}
\slugcomment{Accepted for publication in Astrophysical Journal Letters}
\def\Msun{\hbox{$\rm\thinspace M_{\odot}$}}
\def\spose#1{\hbox to 0pt{#1\hss}}
\def\approxlt{\mathrel{\spose{\lower 3pt\hbox{$\sim$}}
	\raise 2.0pt\hbox{$<$}}}
\def\approxgt{\mathrel{\spose{\lower 3pt\hbox{$\sim$}}
	\raise 2.0pt\hbox{$>$}}}
\mathchardef\twiddle="2218

\begin{document}

\title{Timing Noise Properties of GRO~J0422+32}

\author{ J.E. Grove\altaffilmark{1}, M.S.~Strickman}
\affil{E.O. Hulburt Center for Space Research, Code 7650, Naval
Research Lab., Washington DC 20375, U.S.A.}
\author{S.M.~Matz}
\affil{Northwestern University, Evanston IL 60208}
\author{X.-M. Hua, D. Kazanas, L. Titarchuk}
\affil{NASA/Goddard Space Flight Center, Greenbelt, MD 20771}
\altaffiltext{1}{E-mail:  grove@osse.nrl.navy.mil}

\begin{abstract}
OSSE observed the transient black hole candidate 
GRO~J0422+32 (XN~Per~92) between
1992 August 11 and 1992 September 17.  High time resolution data were obtained
in several energy bands over the $\simeq$35--600 keV range with a timing
resolutions of 8 ms.  Power spectra at energies below 175 keV
show substantial low-frequency red noise with a shoulder at a few $10^{-2}$ Hz,
peaked noise with characteristic frequency near 0.2 Hz, and a second shoulder
at a few Hz.  The frequencies of the shoulders and the peak 
are independent of energy and source intensity.  The complex cross spectrum 
indicates that photons in the 75--175 keV band lag
photons in the 35--60 keV band by a time roughly proportional to the inverse of
the Fourier frequency.  The maximum lag observed is $\simeq$300 ms.  The
power and lag spectra are consistent with the production of the $\gamma$ rays
through thermal Comptonization in an extended hot corona with a power-law
density profile.
\end{abstract}


\section{Introduction}
The hard X-ray transient GRO~J0422+32 (XN~Per~1992) was discovered by the BATSE
instrument on the Compton Gamma Ray Observatory in data from 1992 August 5
(Paciesas et al. 1992), and at its peak reached an intensity in soft $\gamma$
rays approximately three times brighter than the Crab Nebula and pulsar.  The
source was observed by CGRO/OSSE beginning 1992 August 11, 
approximately at the peak of the outburst.

%The results of scanning observations by OSSE were combined with occultation
%studies by BATSE to localize the source in an error circle with radius of
%0.2$^\circ$ (Harmon et al. 1992).  
An optical counterpart was proposed by
Castro-Tirado et al. (1992) and confirmed by the soft $\gamma$-ray observations
of SIGMA (Roques et al. 1994).  
%The region shows no prior history of nova
%outbursts in the Harvard plates (Shao 1992).   
While the mass function
of $1.2 \pm 0.04 \Msun$ determined by Filippenko, Matheson, \& Ho (1995)
is low enough that the compact object might indeed be a neutron star,
the H$\alpha$ radial velocity curve and the M stellar type of the
mass donor imply a mass of 3.6$\Msun$ for the compact primary.  The 
photometric measurements of Callanan et al. (1996) support this mass estimate
and give a distance estimate of $\sim$2 kpc.  

Broadband energy spectra from TTM, HEXE, and OSSE show that during outburst
the source was in the X-ray low, hard state, which coincides
with the breaking $\gamma$-ray
state (Grove et al. 1997, 1998, and references therein).  The gamma radiation
is thus likely the result of thermal Comptonization in a hot corona near
the accretion disk.  The $\gamma$-ray spectrum hardened ($\Delta kT / kT
\simeq +20\%$) as the outburst declined (Grove et al. 1998).

Power spectra above 20 keV show significant red noise and
peaked noise components frequently 
referred to as ``quasi-periodic oscillations'' (QPOs), even though
they do not necessarily satisfy the width requirement (FWHM/$f_0 < $ 0.5)
for such a label.  BATSE
reported ``QPOs'' centered at roughly 0.04 Hz and 0.2 Hz (Kouveliotou et al.
1992), both of which were confirmed by SIGMA (Vikhlinin et al. 1995) and OSSE
(Grove et al. 1992, 1994). 

The spectral shape, rapid variability, and outburst lightcurve are similar to
previous X-ray novae A0620-00 and XN~Mus~1991, both of which have measured mass
functions that make them very strong black hole candidates (BHCs).  Based on 
these similarities, GRO~J0422+32 has been classified as a black hole candidate.

%We report here on timing analyses of the extensive observations of GRO~J0422+32
%with OSSE.  A preliminary report appeared in Grove et al. (1994).

\section{Observations}

The OSSE instrument consists of four identical large-area NaI(Tl)--CsI(Na)
phoswich detector systems (Johnson et al. 1993).  Energy spectra and 
high time-resolution counting rates are accumulated in an alternating 
sequence of two-minute measurements of source and background fields.  High
time-resolution data were collected from the on-source detectors in 8-ms rate
samples in five energy bands from $\simeq$35 keV to $\simeq$600 keV. 
%This
%observation was made in twice-normal gain to improve the low-energy response. 

OSSE observed GRO~J0422+32 on 34 days spanning 1992 August 11 -- 1992 September
17.  The source reached its maximum intensity at 100 keV shortly after the
start of the OSSE pointing, then began a
roughly exponential decline with a decay time
of $\simeq$40 days, falling to about half maximum intensity at 100 keV at the
end of the pointing.

Some 10\% of the total light yield of NaI(Tl) results from a phosphorescence
with a decay time of $\sim$150 ms (McKlveen \& McDowell 
1975).  Fluctuations in this
``afterglow'' from the passage of a heavy cosmic ray can trigger the OSSE
detector system, evidenced as clusters of low-energy
events with a soft spectrum ($\sim$E$^{-5}$, detectable up to $\simeq$100 keV)
and in power spectra of blank sky pointings as weak broad-spectrum noise
roughly consistent with exponential shots with time constant $\simeq$70
ms.  We estimate that the residual noise
power in the 35--60 keV band (normalized to the source intensity)
after a screening process is applied to remove these events is
$<10^{-3}$ (RMS/I)$^2$ Hz$^{-1}$ below 3 Hz and falls as 1/f$^2$ above 3 Hz
(see Fig. \ref{power_spec} for comparison to the noise power from the source).
The residual power is undetectable in the 75--175 keV band.

\subsection{Power Spectrum Analysis}

We obtained the power spectral density in the 35--60 keV and 75--175 keV
energy bands through a multi-step process.  We segmented the data into
two-minute on-source pointings to reduce potential systematic effects that
might arise on long timescales from orbital variations in the background count
rate or differences between source-pointed detectors.  To eliminate the
possibility of spurious power spectral features (i.e. side-lobes) arising from
the window function, we selected only those two-minute pointings that contained
no data gaps or dropouts.  Then we Fourier-transformed 
each 16384-point time series of 8-ms samples, normalized according
to the procedure of Leahy et al. (1983), and subtracted the Poisson noise
contribution, which we corrected for deadtime effects.  We then
summed the power spectra incoherently into daily and longer accumulations. 
Finally, following the prescription of Belloni and Hasinger (1990), we
renormalized the power spectra to the source intensity, which we calculated
from the background-subtracted spectral data from the standard OSSE analysis
(Johnson et al. 1993).  With this normalization, power spectra from different
instruments, sources, and observations are directly comparable.

The normalized power spectral density (PSD) 
$P_k$ at Fourier frequency bin $k$ is given as
the fractional root-mean-square (RMS) variation of the source intensity
$I$, viz.
\begin{equation}
P_k \, df = \frac{N_{tot}} {N_{src}^2} \,
\left ( \frac{2|H_k|^2} {n_{tot}} \, - \, p \right ) \, \Delta f \, ,
\end{equation}
where $N_{tot}$ is the total (i.e. source $+$ background) counts in $M$
Fourier transforms, $N_{src}$ is the source counts---estimated from the
standard spectral analysis---in these $M$ transforms, 
$|H_k|^2$ is the mean Fourier power at frequency $k$,
$p \simeq 2$ is the Poisson noise power after accounting for deadtime effects, 
$n_{tot} = N_{tot} / M$ is the mean of the total (i.e. source $+$ background)
counts per transform, and
$\Delta f = 1/131.072$ Hz is the frequency resolution of the power
spectrum.  We emphasize that this normalization is valid for the 
background-dominated case, which is appropriate for the low-energy gamma-ray
band.  The standard deviation of
the PSD estimator is $ \sigma_k = P_k / \sqrt{M} $
(Bendat \& Piersol 1986), which accounts for both intrinsic source
noise and Poisson noise.

Figure \ref{power_spec}a shows the normalized PSD in the
35--60 keV and 75--175 keV bands for the entire OSSE pointing.  The total
fractional RMS variation between 0.01 Hz and 60 Hz is $\simeq$40\% in 35--60
keV, and $\simeq$30\% in 75--175 keV.  The shape of the power spectrum is
essentially identical in the two energy bands.  It shows breaks at a few times
$10^{-2}$ Hz and a few Hz, and a strong peaked-noise component (frequently
labelled a ``QPO'') at 0.23 Hz, with FWHM $\simeq 0.2$ Hz.  Statistically
significant red noise is detected at frequencies up to $\sim$20 Hz.  Not
readily apparent in this figure is an intermittent peaked noise component at
about 0.04 Hz.  The amplitude of this feature varies from day to day.  The two
peaked-noise components and the lower-frequency spectral break have been
reported elsewhere (Kouveliotou et al. 1992, Grove et al. 1992, Denis
et al. 1994).

OSSE's high sensitivity and high sampling rate
have made the second spectral break apparent and permitted a study of the
evolution of the various components that comprise the noise spectrum.  We 
found (Grove et al. 1994) 
that the integrated 0.01--60 Hz power increased as the intensity of the source
decreased, i.e. total power was anticorrelated with intensity and 
correlated with spectral
hardness.  In addition, while the intensity in these energy bands
dropped by nearly a factor of two and the energy spectrum hardened by
$\sim$20\% in effective temperature over the course
of the OSSE observation, the frequencies of the two breaks and the main peaked
noise remained constant, i.e. there was no evidence for significant 
variability in the timescales of the noise processes.
%, which
%contrasts with the ``accordion-like'' stretching of
%the power spectrum of GRO~J1719--24 as its discovery outburst evolved
%(van der Hooft et al. 1996), and with the variability of the power spectral
%break frequencies in Cyg~X-1 in the X-ray low, hard state (e.g. Belloni
%\& Hasinger 1990).  

\subsection{Lag Spectrum Analysis}

From the cross-spectral density (Bendat \& Piersol
1986), one can measure the phase or time lag between two series as a
function of Fourier frequency.  Given two time series, e.g. of soft photons
$s(t_k)$ and hard photons $h(t_k)$, the cross-spectral density $C_{sh}(f_k)$ is
given by 
\begin{equation}
C_{sh}(f_k) df = \frac{2}{M} \sum{S_m^*(f_k) H_m(f_k)} \Delta f
\end{equation}
where the sum runs over the $M$ Fourier transforms $S_m(f_k)$ and $H_m(f_k)$
of the segmented soft and hard time series, respectively.  The phase difference 
$\Delta \phi_{sh}(f_k)$ at Fourier frequency $f_k$ between the two series is 
then
\begin{equation}
\Delta \phi_{sh}(f_k) = \arctan \left( \frac{Im[C_{sh}(f_k)]}{Re[C_{sh}(f_k)]}
\right)
\end{equation}
where $Im[C_{sh}(f_k)]$ and $Re[C_{sh}(f_k)]$ are the imaginary and real parts,
respectively, of the cross-spectral density.  
%If the sign of the imaginary part
%is positive at frequency $f_k$, the soft time series leads the hard time
%series; if the imaginary part is negative, the soft lags the hard.  
Time lags 
are simply computed from the phase lags by dividing by $2 \pi f_k$.  Following
van der Klis et al. (1987), we correct for deadtime-induced cross-talk
between the two bands by subtracting the mean cross-spectral density in the
40--62.5 Hz range from the entire cross spectum.  Because the imaginary part is
negligible in this frequency range, the subtraction does not alter the sign of
the phase differences at lower frequencies, and it has negligible
effect on the amplitude of the phase differences at frequencies below
$\sim$10--20 Hz.  We have used the standard deviation of the phase difference
given by Bendat \& Piersol (1986), 
%, where the spectra are substantially influenced, and are
%eventually dominated, by Poisson noise.  The standard deviation 
%of the phase difference is (Bendat \& Piersol 1986)
%\begin{equation}
%\delta \Delta \phi_{sh}(f_k) = 
%\frac{\sqrt{(1 - \gamma_{sh}^2 (f_k))}}{|\gamma_{sh}(f_k)|} 
%\frac{1}{\sqrt{2M}} \, ,
%\end{equation}
%where the coherence function $\gamma_{sh}^2$ is given by
%\begin{equation}
%\gamma_{sh}^2 (f_k) = \frac{|C_{sh}(f_k)|^2} {|S(f_k)|^2 |H(f_k)|^2} \, .
%\end{equation}
which is strictly applicable only for noiseless measurements.  Poisson noise 
causes this to be an underestimate above a few Hz (Vaughan
\& Nowak 1997), but this effect does not alter our conclusions.

The phase and time delay spectra we derive for the entire observation of
GRO~J0422+32 are shown in Fig. \ref{lag_spec}.  The hard emission (75--175 keV)
lags the soft emission (35--60 keV) at all Fourier frequencies, except above 10
Hz, where there is no statistically significant lag or lead between the two
bands.  The phase lag is a weak function of Fourier frequency and
peaks near 1 Hz.  The peak lies far below the Nyquist frequency and is
therefore not a consequence of the finite data binning, as discussed
by Crary et al. (1998).  At frequencies
$\sim$0.01 Hz, hard lags as large as 300 ms are observed, and the time lag
falls roughly as $1/f$.  There is no significant change in the lag at the
frequencies dominated by the strong peaked noise component at 0.23 Hz. 

%The sensitivity of the cross-spectral analysis is such that
%lags are detectable on a daily basis.  We searched for variability in
%the phase lags in several broad frequency bands, including the band
%0.2--1.0 Hz dominated by the peaked noise.  We find weak evidence that the hard
%phase lag in the $\sim$0.3--2.0 Hz range decreases as the source luminosity
%decreases, while in other frequency ranges the lag remains roughly constant.

\section{Discussion}

Generally similar power spectra have been reported from a number of black hole
candidates, and beginning with Terrel (1972), they have frequently been modeled
as arising from a superposition of randomly occurring bursts, or ``shots''.  If
the shots have an instantaneous rise and exponential decay (or vice versa) with
time constant $\tau$, the resulting power spectrum is constant below the
characteristic frequency $ 1 / (2 \pi \tau )$ and falls as $ 1 / f^2 $ at high
frequencies.  This type of model can describe the two breaks and the $ 1/ f^2$
behavior above several Hz in the power spectrum of GRO~J0422+32 if there
exist (at least) two independent shot components, with e-folding times 
$\tau_s \simeq 50$ ms and $\tau_l \simeq 2.1$ sec.  The PSD of
the two-shot model is shown for the 75--175 keV band in 
Fig. \ref{power_spec}a.  
%Note that it
%cannot be that one shot timescale modulates the other, e.g. random 50-ms shots
%with amplitudes following a 2.1-sec exponential envelope, because the power
%spectrum of such a convolution is the product, rather than the sum, of the
%power spectra of the components.  Note also that the best-fit values of
Note that the best-fit values of
the long and short e-folding times and the ratio of amplitudes of the two
components are independent of energy (Table \ref{shot_fit}).

Subtracting the PSD of the two-shot model from the observed PSD gives a 
peaked noise profile that is broad and asymmetric, with a sharp low-frequency 
edge and a broad high-frequency tail, as shown in 
Fig. \ref{power_spec}b.  Plausible alternative descriptions of the continuum
between 0.1 and 1.0 Hz, e.g. a simple power law with index $-0.9$, do not
significantly alter the shape of the peaked noise, although they may change its
amplitude.  The sharp low-frequency edge indicates that the physical process
responsible for the peaked noise has a well-defined maximum timescale.  This
process may perhaps be thermal-viscous instabilities in the accretion disk
(Chen \& Taam 1994) or oscillations in a Comptonizing corona (Cui et al. 
1997).

We attempted to fit the total PSD with simple analytic forms---e.g. in
the time domain, multiple exponentially-damped sinusoids; or in the
frequency domain, multiple
zero-centered Lorentzians to model the continuum and offset Lorentzians to 
model the peaked noise---but none of these adequately describes the
sharp rise and broad fall of the peaked noise, nor do they add significantly
to our understanding of the characteristic timescales represented in the PSD.
Similarly, the scenario of Vikhlinin, Churazov, \& Gilfanov (1994), in
which shots arise from a common reservoir and are coupled through a weak 
amplitude or probability interaction that generates QPOs, also fails to 
describe the observed PSD in detail.

The lag spectrum (Fig. \ref{lag_spec}) is generally similar to that of
several other BHCs in the Ginga or Rossi XTE/PCA band 
(i.e. below 40 keV).  In the X-ray low, hard state, these include
Cyg~X-1, GX339--4, and GS2023+338 (Miyamoto et al. 1992), and
1E1740.7--2942 and GRS~1758--258 (Smith et al. 1997).
In the X-ray very high state, BHCs with similar lag spectra 
are GS1124--683 and GX339--4, subtype ``C+D''
for the latter object (Miyamoto et al. 1993).  Furthermore, the lag
spectrum is quite
similar to that between 20-50 keV and 50-100 keV from Cyg~X-1,
which appeared to be essentially independent of the X-ray or $\gamma$-ray 
state (Crary et al. 1998).  The present result is
more evidence indicating that the frequency-dependent time lag
is a common phenomenon shared by many accreting objects
in binaries.

The observed power and lag spectra are at odds with the predictions
of accretion models that produce most of the X-ray and $\gamma$-ray
emission from a region whose size is comparable to that of the last stable
orbit around a black hole of mass a few M$_{\odot}$.  The characteristic
time scale associated with the dynamics of accretion in such an object
is of order $10^{-3}$ sec; hence one would expect most of
the associated power in the kHz frequency range.  By contrast there is a
remarkable {\it lack} of power at this range.  Furthermore, under 
these conditions the time lags, which in these models are indicative of the
photon scattering time in the hot electron cloud, should be independent
of the Fourier frequency and also of order $10^{-3} $ sec, the photon
scattering time in this region.

Miller (1995) has argued that the observed time lags represent lags instrinsic
to the soft seed photons, rather than the Comptonizing cloud.  However,
Nowak \& Vaughan (1996) have shown that any intrinsic lag is washed out
if the observed photon energies are much greater than the seed photon energies,
as is the case here, leaving again a frequency-independent lag due to the
difference in scattering times across the cloud.

The discrepancy between observed and predicted power and lag spectra
prompted an alternative approach proposed
recently by Kazanas, Hua \& Titarchuk (1997; hereafter KHT) and Hua, 
Kazanas \& Titarchuk (1997).  These authors suggested
that, while the process  responsible for the formation of the 
high energy spectra is indeed Comptonization, the hot, scattering electron 
cloud extends over several decades in radius with a power law profile in 
density, $n(r) \propto 1/r^p$.
% In particular, they considered a density profile of the form
%\begin{equation}
%n(r) = \left\{ \begin{array}{ll}
%    n_i	             &  \mbox{for $r \le r_1$} \\
%    n_1 (r_1/r)^{p}  &  \mbox{for $r_2 > r > r_1$}
%\end{array}
%\right.
%\end{equation}
%where $p$ is a free parameter; $r$ is the radial distance from
%the center of a spherical cloud; $r_1$ and $r_2$ are radii of the
%inner and outer edges of the atmosphere respectively. The central core
%is assumed to be uniform and its density is $n_i$, which may not equal
%to $n_1$.
This power-law density profile has a number of properties of interest in 
interpreting timing and spectral observations.

For a $\delta-$function injection of soft photons at
the center of the cloud, the light curves
of the photons  emerging from the cloud at a given energy are 
power laws extending in time to $\sim r/c$ ($r$ is the outer 
edge radius of the atmosphere) followed by an exponential cutoff. For 
small values of the total Thomson depth $\tau_0$, the power-law 
index of the light curve is roughly equal to the power-law 
index $p$ of the density profile of the 
scattering cloud, becoming progressively flatter for increasing values
of $\tau_0$ and higher escaping energies (Fig. 1 in KHT). On the 
other hand, the corresponding light curves for clouds of uniform 
density are exponentials without power-law portions.  The
time dependence of the photon flux can therefore be used to map the radial
density profile  of the scattering cloud.
%  This fact provides 
%the possibility of deconvolution of the density 
%structure of Comptonizing clouds through analysis of their light curves.

%The form of the light curves has direct bearing 
%on the resulting PSD, since for an ensemble of uncorrelated shots with 
%fixed profiles, injected at a given constant rate 
%and at Poisson-distributed intervals, the PSD can be obtained by 
%computing the same quantity for a single shot. 
%The light curves (the response of the cloud to a $\delta$-function
%injection of soft photons, such as those in Figs. 1 and 2 of
%KHT) can be approximated by a power-law times exponential form.
%Thus, the specific form and the size of the scattering 
%atmosphere provide an account of the characteristic low frequency turn over of
%the PSD, a feature absent in the other models purporting to account for the 
%timing properties of X-ray binaries.

For a uniform cloud, the density profile has index
$p=0$, and the light curve has no power law portion, 
i.e. the resulting PSD is that corresponding to an exponential
shot.  For a density profile with index $p=1$ and total Thomson depths
in the scattering atmosphere of a few, the PSD is 
$\propto 1/f$ (KHT Fig. 1).  One should note that this
form of the PSD assumes infinitely sharp turn-on of the shots at $t=0$.  As 
Kazanas \& Hua (1997) have shown, a finite turn-on time $t_0$ will 
introduce an additional steepening of 
the PSD at frequencies $\omega \sim 1/t_0$ extending over a
decade in frequency, yielding PSDs in agreement with those of 
Fig. \ref{power_spec}a.  The great advantage of the 
present scheme is therefore the direct
physical association of features in the PSD with properties of the source. 
Modeling of the light curves of GRO~J0422+32 with this type of shot indicates 
values for $t_0 \approx 50$ msec.

The model presented
in KHT provides constraints on the time lags that can be of great value
in probing the structure of the scattering medium.  In the process
of Comptonization, photons of energy $E_2$ lag in time behind photons of energy
$E_1 < E_2$ simply because more scatterings are required to take a photon 
from $E_1$ to $E_2$.  The lag in time is proportional to the scattering
time, which depends only on the density of the medium. Thus in general, for a 
uniform medium the lag time is constant (i.e. independent of the Fourier 
period).  However, in a medium with a power-law density profile, the hard
photons sample a range of several orders of magnitude in density, which 
appears in the corresponding time lags.  In addition, because 
the probability of scattering at a given density range
is constant for a medium with $p=1$, all lags should be present with 
equal weight, producing a time-lag function $\propto 1/f$, with
a maximum lag at the time scale
corresponding to the scattering time at the edge of the power-law 
atmosphere.  Indeed, Fig. \ref{lag_spec} 
is in excellent agreement with the above arguments (see Hua, Kazanas, \&
Cui 1997a for fits to similar lag spectra from Cyg~X-1, and Hua, Kazanas, \&
Cui 1997b for discussion regarding preliminary OSSE data from GRO~J0422+32).
%Interestingly, the time lags level off at the same timescales at which the 
%PSD do, thus providing supporting evidence for the power-law shot model 
%discussed in KHT and HKT. 

\acknowledgements
This work was supported under NASA DPR S-10987C.

\clearpage

\begin{table}
\begin{center}
\begin{tabular}{lcc}
\tableline
\tableline
Parameter	&	35--60 keV	&	75--175 keV\\
\tableline
& \\
$\tau_l$	&       $2.1 \pm 0.1$ s	&	$2.1 \pm 0.1$ s\\
$\tau_s$	&       $53 \pm 3$ ms	&	$51 \pm 2$ ms\\
$A_l / A_s$\tablenotemark{a} & $2.0 \pm 0.2$	&	$1.9 \pm 0.1$\\
\tableline
\end{tabular}
\end{center}
\caption{Exponential shot fits to power spectra.}
\tablenotetext{a}{Ratio of amplitude of long to short decay-time 
shot components.}
\label{shot_fit}
\end{table}

\clearpage

\begin{figure}
\caption{(a) Normalized power spectra for GRO~J0422+32 in 35--60 keV (upper
curve, diamonds) and 75--175 keV (lower curve, crosses) bands.  Model 
fit (solid lines) to the 75--175 keV band includes exponential shots with 
lifetimes $\simeq$50 ms and $\simeq$2.1 sec.  (b)  Residual 
power after two-shot model is subtracted.  Peaked noise components are visible
near 0.04 Hz and 0.23 Hz.  For clarity, only the 75--175 keV band is shown.} 
\label{power_spec}
\end{figure}

\begin{figure}
\caption{(a) Phase and (b) time lag of the hard emission (75--175 keV) relative
to the soft (35--60 keV) emission from GRO~J0422+32 as a function of Fourier
frequency.}
\label{lag_spec}
\end{figure}

\begin{figure}
\vbox{\psfig{figure=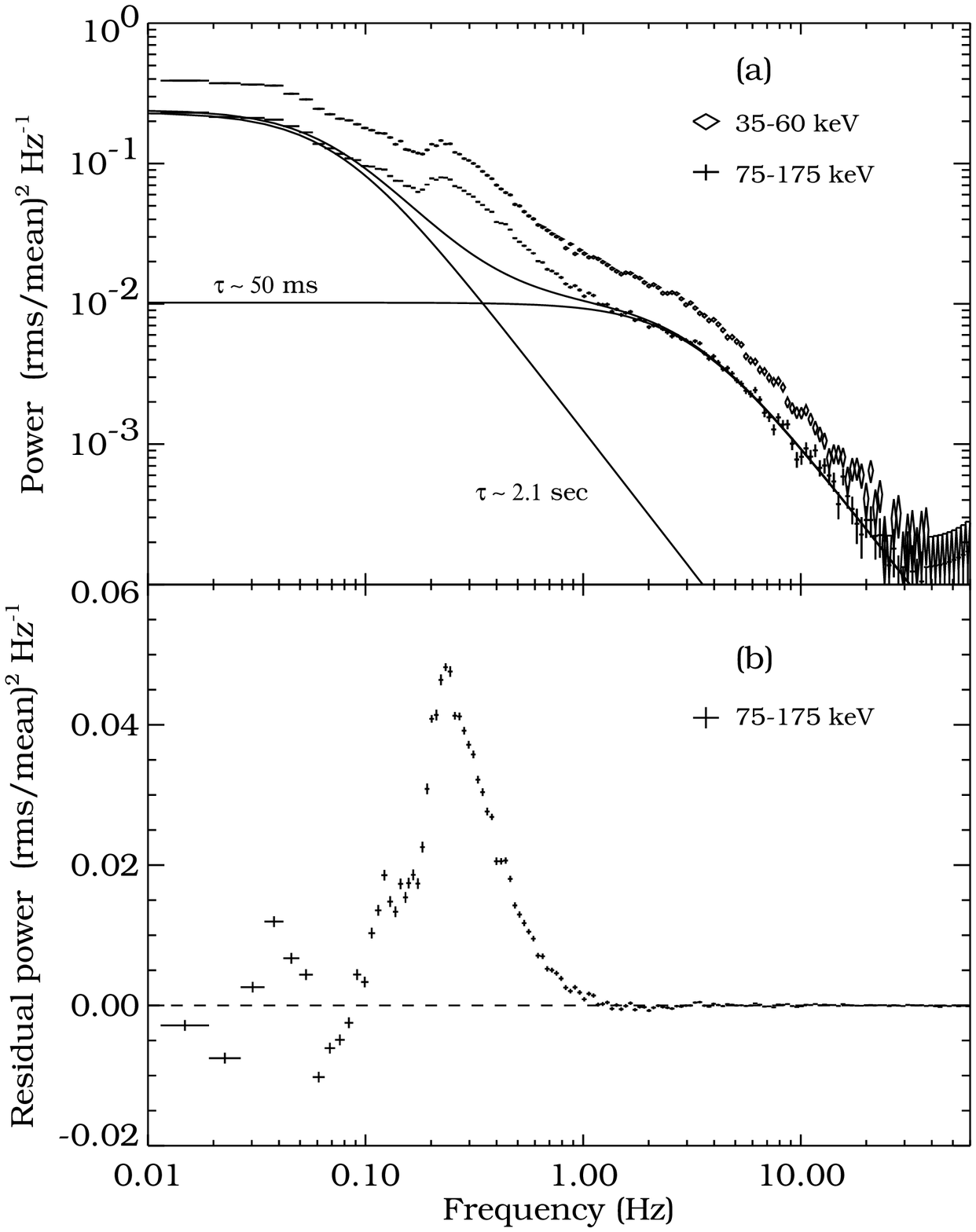}}
\end{figure}

\begin{figure}
\vbox{\psfig{figure=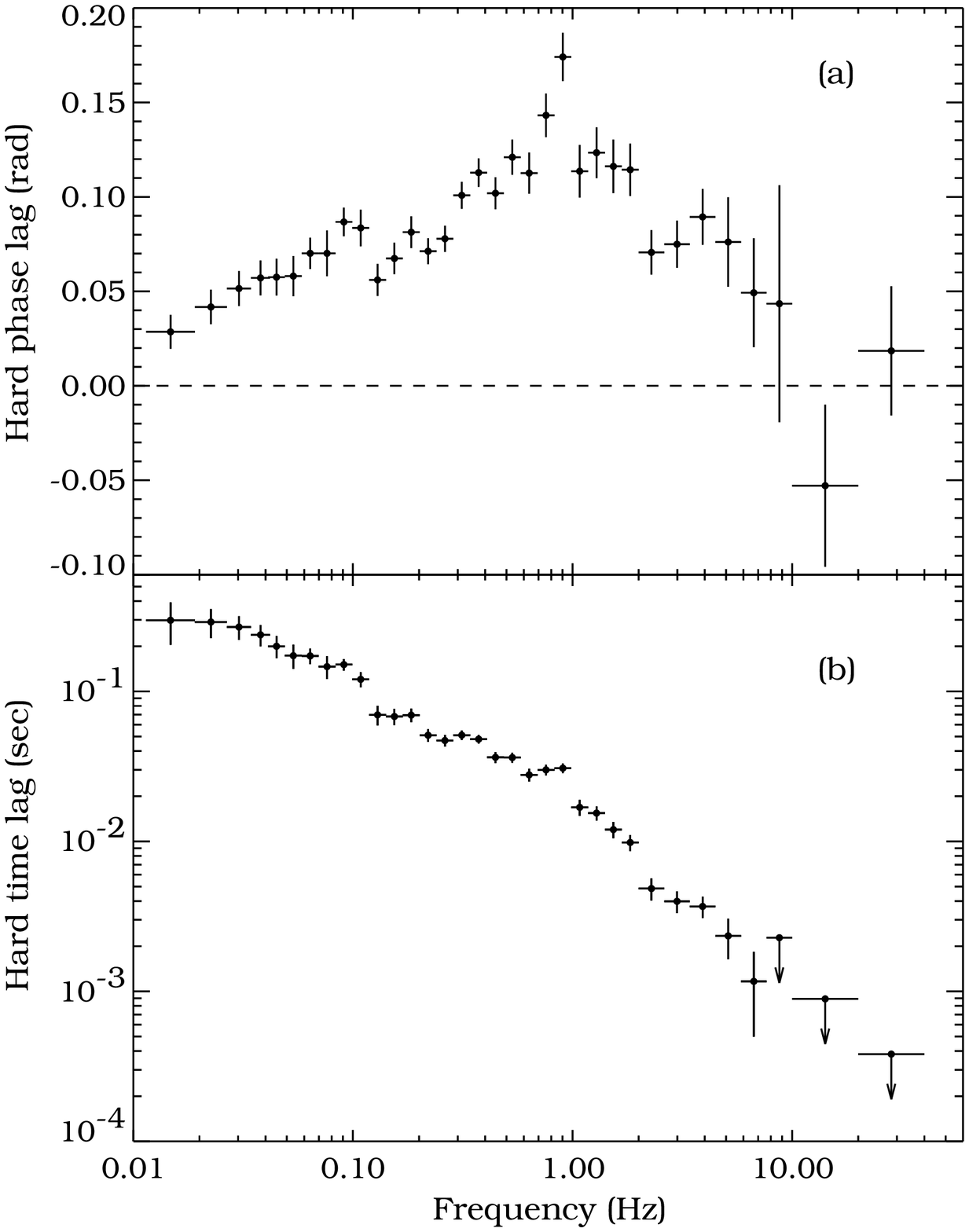}}
\end{figure}

\end{document}